\documentclass[aps,twocolumn,prl,showpacs,amsmath]{revtex4}
\usepackage[dvips]{graphicx}

\begin{document}

\title{Virial Theorem and Universality in a Unitary Fermi Gas}
\author{J. E. Thomas, J. Kinast and A. Turlapov}
\affiliation{Physics Department, Duke University, Durham, North
Carolina 27708-0305} 

\date{\today}

\begin{abstract}
Unitary Fermi gases, where the scattering length is large compared
to the interparticle spacing,  can have universal properties,
which are independent of the details of the interparticle
interactions when the range of the scattering potential is
negligible. We prepare an optically-trapped, unitary Fermi gas of
$^6$Li, tuned just above the center of a broad Feshbach resonance.
In agreement with the universal hypothesis, we observe that this
strongly-interacting many-body system obeys the virial theorem for
an ideal gas over a wide range of temperatures. Based on this
result, we  suggest a simple volume thermometry method for unitary
gases. We also show that the observed breathing mode frequency,
which is close to the unitary hydrodynamic value over a wide range
of temperature, is consistent with a  universal hydrodynamic gas
with nearly isentropic dynamics.


\end{abstract}

\maketitle


Universal behavior is believed to approximately describe a variety
of strongly-interacting Fermi systems, such as neutron
stars~\cite{Heiselberg,HeiselbergNeutronStar,Bertsch,Baker} and
resonantly interacting atomic Fermi gases~\cite{Heiselberg}.
Universal Fermi systems satisfy the unitary condition, where  the
zero energy scattering length greatly exceeds the interparticle
spacing, while the range of the scattering potential is
negligible. Unitary conditions are produced in an optically
trapped Fermi gas~\cite{AmScientist}, by using a magnetic field to
tune near a broad Feshbach resonance, where strong interactions
are observed~\cite{OHaraScience}. According to the universal
hypothesis, the interparticle spacing sets the only natural length
scale, and the system can exhibit universal
thermodynamics~\cite{Heiselberg,Bertsch,Baker,OHaraScience,HoHighTemp,HoUniversalThermo,BruunUniversal}.
 If unitary Fermi gases satisfy this hypothesis, they can be
used to test predictions  in fields well outside of atomic
physics.

Universality at low temperature has been tested in  measurements
of two-component, strongly-interacting Fermi gases. The ratio,
$\beta$, of the interaction energy to the local Fermi energy has
been
measured~\cite{OHaraScience,MechStab,SalomonInteraction,Grimmbeta,JointScience}
and is believed to be a zero-temperature, universal
parameter~\cite{Heiselberg}. The measurements are in reasonable
agreement with recent predictions~\cite{Carlson,Strinati}.
Universality also may explain the small line shifts observed in
radio-frequency spectroscopy measurements~\cite{GuptaRF}.
Universality at low temperature is further supported by the
spatial profile~\cite{OHaraScience,Grimmbeta,JointScience} and
breathing mode frequency~\cite{Kinast,KinastMagDep,KinastDampTemp}
of a trapped, highly degenerate, unitary Fermi gas. However, for a
Fermi gas tuned to a broad Feshbach resonance, there have been no
model-independent tests of the universal hypothesis over a wide
temperature range, and these tests impact all theoretical
predictions of the thermodynamics.

In this Letter, we show theoretically and  experimentally that a
harmonically trapped, unitary $^6$Li Fermi gas obeys the virial
theorem for an ideal gas over a wide range of temperature, as
predicted by universal thermodynamics. Although the gas is a
strongly-interacting, many-body system, we find that mean square
width of the cloud varies linearly with the total energy, as
predicted by the virial theorem. We also show that universal
hydrodynamics, under isentropic conditions, may describe  the
behavior of the observed breathing mode
frequency~\cite{KinastDampTemp}, which is near the zero
temperature unitary hydrodynamic value over a wide range of
temperatures, and has no mean field shift.

We test the universal hypothesis in a highly degenerate, unitary
Fermi gas of $^6$Li, confined in a CO$_2$ laser trap. This
far-detuned trap provides a nearly harmonic potential, which is
the same for all atoms, both paired and unpaired. We employ a
50-50 mixture of the two lowest spin-up and spin-down hyperfine
states, and tune a bias magnetic field to 840 G, just above the
center of a broad s-wave Feshbach resonance. Forced evaporation in
the optical trap is then used to cool the gas, as described
previously~\cite{OHaraScience,Kinast,KinastMagDep,KinastDampTemp,JointScience}.
After evaporation, we obtain a total of $N=2.0(0.2)\times 10^5$
atoms, and the cloud has a nearly zero-temperature Thomas-Fermi
profile, as expected for a unitary gas at zero
temperature~\cite{JointScience}.

From the measured trap frequencies  we obtain
$\omega_\perp=\sqrt{\omega_x\omega_y} = 2\pi\times 1696(10)$ Hz,
$\omega_x/\omega_y=1.107(0.004)$, and $\lambda
=\omega_z/\omega_\perp=0.045$. The typical Fermi temperature (at
the trap center for a noninteracting gas) is $T_F=(3
N)^{1/3}\hbar(\omega_x\omega_y\omega_z)^{1/3}/k_B\simeq
2.4\,\mu$K, small compared to the final trap depth of
$U_0/k_B=35\,\mu$K. The coupling parameter of the
strongly-interacting gas at $B=840$ G is $k_Fa\simeq -30.0$, where
$\hbar k_F=\sqrt{2m\, k_B T_F}$ is the Fermi momentum, and
$a=a(B)$ is the zero-energy scattering length estimated from the
measurements of Bartenstein et al.,~\cite{BartensteinFeshbach}.

We now show that universality requires such  a
strongly-interacting, unitary Fermi gas to obey the virial theorem
for a harmonically-trapped ideal gas at all temperatures.
According to the universal hypothesis, the thermodynamic functions
which describe the gas must be independent of the interaction
parameters and can depend only on the total density $n$ and
temperature $T$~\cite{HoUniversalThermo}.

Consider first the local energy $\Delta E$ (kinetic and
interaction energy) contained in a small volume $\Delta V$ of  gas
centered at position $\mathbf{x}$ in a harmonic trap. Assume that
the volume $\Delta V$ contains a fixed number of atoms $\Delta N$,
so that $n=\Delta N/\Delta V$, where $\int
d^3\mathbf{x}\,n(\mathbf{x})=N$ is the total number of trapped
atoms.

For such a unitary gas,  the local energy  must be of the general
form,
\begin{equation}
\Delta E = \Delta N\,
\epsilon_F(n)\,f_E\left[\frac{T}{T_F(n)}\right].
\label{eq:localenergy}
\end{equation}
Here, the natural energy scale for atoms of  mass $m$ is taken to
be $k_BT_F(n)=\epsilon_F(n)$ with $\epsilon_F(n)\equiv
\hbar^2\,(3\pi^2\,n)^{2/3}/(2m)$. With this definition,
$\epsilon_F(n)$ is the local Fermi energy corresponding to the
density $n$ and $T_F(n)$ is the corresponding local Fermi
temperature. Note that for a zero-temperature ideal Fermi gas, we
have $f_E= 3/5$, for a zero-temperature unitary gas
$f_E=3(1+\beta)/5$, while for a classical gas, $f_E =
(3/2)T/T_F(n)$.

The corresponding local entropy $\Delta S$ takes the form
\begin{equation}
\Delta S = \Delta N\, k_B\,f_S\left[\frac{T}{T_F(n)}\right],
\label{eq:localentropy}
\end{equation}
where $k_B\,f_S$ is the average entropy per particle, which can
contain normal and superfluid contributions.

The local pressure of the gas is readily determined from the
relation $P=-[\partial (\Delta E)/\partial (\Delta V)]_{\Delta
N,\Delta S}$. From Eq.~\ref{eq:localentropy}, we see that holding
the local entropy constant requires $f_S\,=$ constant, which in
turn means that we hold the local reduced temperature constant in
taking the derivative of $\Delta E$ with respect to volume $\Delta
V$. Hence, we need only to find the volume derivative of the local
Fermi energy, which  yields the local pressure,
\begin{equation}
P = \frac{2}{3}\, {\cal E}(n,T) , \label{eq:pressure2}
\end{equation}
where the local energy density (total kinetic plus interaction
energy per unit volume) is ${\cal
E}(n,T)=n\,\epsilon_F(n)\,f_E[T/T_F(n)]$. Eq.~\ref{eq:pressure2}
relates the pressure and local energy density for the unitary gas
in the same way as for an ideal, noninteracting homogeneous gas,
although the energy densities are quite different.
Eq.~\ref{eq:pressure2}  for a unitary gas was obtained previously
in Ref.~\cite{HoUniversalThermo}.

 In mechanical equilibrium, the balance of the forces arising from
 the pressure $P$ and trapping potential $U$ yields
\begin{equation}
\nabla P(\mathbf{x})+n(\mathbf{x})\,\nabla U(\mathbf{x})=0.
\label{eq:forcebal2}
\end{equation}
 Taking an inner product of
Eq.~\ref{eq:forcebal2} with $\mathbf{x}$ and using
\mbox{$\mathbf{x}\cdot\nabla U(\mathbf{x})=2\,U(\mathbf{x})$} for
a harmonic trap, one readily obtains $N\langle U\rangle =
(3/2)\int d^3\mathbf{x}\, P(\mathbf{x})$, where $\langle U\rangle$
is the average potential energy per particle. Using $\int
d^3\mathbf{x}\, {\cal E}(\mathbf{x})=E-N\langle U\rangle$ and
Eq.~\ref{eq:pressure2} then yields
\begin{equation}
N\langle U\rangle=\frac{E}{2}.
\label{eq:virial}
\end{equation}
Hence, universality requires a unitary Fermi gas  to obey the
virial theorem for an ideal gas. Since the mean square size is
$\propto\langle U\rangle$, Eq.~\ref{eq:virial} is equivalent to
\begin{equation}
\frac{\langle x^2(E)\rangle}{\langle
x^2(E_0)\rangle}=\frac{E}{E_0}\, , \label{eq:meansq}
\end{equation}
 which we use to verify the theorem. Here $E_0$ is the ground state energy of the
 cloud.

Energy is first added to gas, always starting from the lowest
temperatures, by abruptly releasing the cloud and then recapturing
it after a short expansion time $t_{heat}$~\cite{JointScience}.
During the expansion time, the total kinetic and interaction
energy  is conserved. When the trapping potential $U(\mathbf{x})$
is reinstated, the potential energy of the expanded gas is larger
than that of the initially trapped gas, increasing the total
energy to $E(t_{heat})$, which is a known function of
$t_{heat}$~\cite{JointScience}. After waiting for the cloud to
reach equilibrium, the sample is released from the trap. The mean
square width $\langle x^2\rangle$ is estimated by fitting a
one-dimensional, finite-temperature, Thomas-Fermi profile to the
spatial distribution of the cloud~\cite{JointScience}, which is
imaged after a {\it fixed} expansion time of 1 ms.

\begin{figure}
    \includegraphics[width=3.0in]{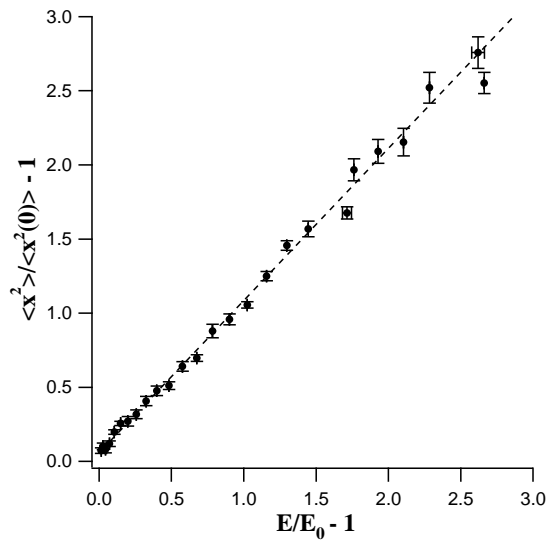}
     \caption{Verifying the virial theorem in a unitary Fermi gas of $^6$Li: $\langle x^2\rangle/\langle x^2(0)\rangle$ versus $E/E_0$
 showing linear scaling. Here $\langle
x^2\rangle$ is the measured transverse mean square size. $E$ is
the total energy, calculated as in Ref.~\cite{JointScience}. $E_0$
and $\langle x^2(0)\rangle$ denote ground state values.
\label{fig:Evssize}}
\end{figure}

Fig.~\ref{fig:Evssize} shows $\langle x^2\rangle$  as a function
 $E$. The dashed line shows the fit, $\langle x^2\rangle/\langle
x^2(0)\rangle  = 1.03\, (0.02)\,E/E_0$, which is in close
agreement with the virial theorem prediction of
Eq.~\ref{eq:meansq} for a unitary gas. Eq.~\ref{eq:virial} is
therefore verified because $N\langle U(T=0)\rangle = E_0/2$, which
follows generally from the equation of state for a unitary gas at
zero temperature~\cite{OHaraScience,JointScience,MechStab}. The
Fermi radius $\sigma_x'$ of the trapped unitary cloud is measured
from a fit at nearly zero temperature~\cite{JointScience}, and
determines $\langle x^2(T=0)\rangle=\sigma_x'^2/8$ as well as
$E_0/N=3m\omega_x^2\langle x^2(T=0)\rangle$.

From these results, we see that despite the strong, many-body
interactions, the total potential energy of the unitary gas is
half of the total energy. Hence, the sum of the kinetic and
interaction energies must be exactly half of the total energy at
all temperatures.

We suggest that an empirical thermometry method can be based on
the virial theorem result for a unitary Fermi gas. One can simply
measure $\langle x^2\rangle$ for a unitary gas near a broad
Feshbach resonance. This determines the total energy according to
Eq.~\ref{eq:meansq}, i.e., since $E/N=3m\omega_x^2\langle
x^2\rangle$. Universality then requires that $E/E_0$ is in
one-to-one correspondence with the reduced temperature, $T/T_F$.

An approximate empirical reduced temperature  can be determined by
assuming that $E/E_0$ obeys ideal gas scaling with reduced
temperature. At a later time, the empirical temperature can be
calibrated, either theoretically or experimentally.

A theoretical calibration, relating the empirical and  theoretical
reduced temperatures,  can be accomplished using an exact
calculation of $E(T/T_F)/E_0$ for a unitary gas. While this
calibration method is not useful for measurements of the energy
versus temperature (where it is a tautology), it can be used to
estimate temperature for precise quantitative comparisons between
predictions and measurements of  condensed pair
fractions~\cite{Jincondpairs,Ketterlecondpairs}, the
gap~\cite{GrimmGap}, collective mode damping
rates~\cite{Kinast,KinastDampTemp}, etc.

An experimental consistency check of the calibration can be done
by measuring the total entropy of the unitary gas versus energy,
$S(E)$. $E$ is known from $\langle x^2\rangle$. $S$ can be
determined, albeit in a model-dependent way at
present~\cite{JointScience,CastinSweep}, using an adiabatic sweep
of the magnetic field~\cite{CastinSweep} from the unitary regime
to a weakly interacting regime either well
below~\cite{GrimmGap,Ketterlecondpairs} or well above
resonance~\cite{Jincondpairs}, where $S$ is
calculated~\cite{JointScience,CastinSweep}. Then, the fundamental
relation, $1/T=(\partial S/\partial E)$ yields the temperature of
the unitary gas for a given $E$, which can be compared to the
empirical temperature measured at the corresponding $E$.


 We now consider the implications of universal
  hydrodynamics under   isentropic expansion conditions, which may apply
  to the radial breathing mode of a unitary Fermi gas~\cite{KinastDampTemp}.
  We find that  the damping rate reveals a transition in behavior,
  while the frequency remains
close to the zero-temperature unitary hydrodynamic value over a
wide range of temperature~\cite{KinastDampTemp}.

Under locally isentropic conditions,  the stream velocities of the
normal and superfluid components must be equal (since the entropy
per particle is different for the superfluid and normal
components). Then, we can   assume that the  total density
$n(\mathbf{x},t)$ and the stream velocity
$\mathbf{u}(\mathbf{x},t)$ obey
 a simple hydrodynamic equation of motion.
 The convective derivative of the stream velocity
$\mathbf{u}$ is the local acceleration which depends on the forces
arising from the local pressure $P$ and the trap potential $U$.
For irrotational flow, $\nabla\times\mathbf{u}=0$, we have
$\mathbf{u}\cdot\nabla\mathbf{u}=\nabla (\mathbf{u}^2/2)$. Then,
\begin{equation}
m\frac{\partial\mathbf{u}}{\partial t}=-\nabla
\left(\frac{m}{2}\mathbf{u}^2+U\right)-\frac{\nabla P}{n},
\label{eq:hydro}
\end{equation}
where $m$ is the atom mass~\cite{baremass}.

 For a unitary Fermi gas, the local pressure, Eq.~\ref{eq:pressure2} takes the general form
$P(n,T)=n^{5/3}\,f_P[T/T_F(n)]$.

 Initially, the gas is contained in a harmonic trap at a uniform
 temperature $T_0$ and has a density $n_0\equiv n_0(\mathbf{\tilde{x}})$, where $\mathbf{\tilde{x}}$
is the position in the initial distribution. The initial pressure
$ P_0(\mathbf{\tilde{x}})=n_0^{5/3}\,f_P[T_0/T_F(n_0)]$. Force
balance, Eq.~\ref{eq:forcebal2}, requires
$\nabla_{\mathbf{\tilde{x}}}P_0(\mathbf{\tilde{x}})/n_0=-\nabla_{\mathbf{\tilde{x}}}U(\mathbf{\tilde{x}})$.

The hydrodynamic equation of motion can be solved by assuming a
scaling ansatz~\cite{Menotti,Guery}, where each dimension changes
by a scale factor $b_i(t)$, $i=x,y,z$, and $b_i(0)=1$. The density
and stream velocity then take the forms
$n(\mathbf{x},t)=n_0(\mathbf{\tilde{x}})/\Gamma$ and
$u_x=x\,\dot{b}_x(t)/b_x(t)$
 and similarly for $u_y,u_z$. Here $\mathbf{\tilde{x}}\equiv (x/b_x,y/b_y,z/b_z)$
 is the position  at time $t=0$ for an atom which is at position $\mathbf{x}$ at
 time $t$ and $\Gamma\equiv b_xb_yb_z$ is the
 volume scale factor.
 The scaling ansatz is exact if the gas is contained in a harmonic
trap and the pressure takes the form $P=c\,n^\gamma$, where $c$
and $\gamma$ are constants~\cite{Menotti,Guery}.

At nonzero temperature, the pressure does not in general obey such
a simple power law, as the function $f_P$ can be dependent on
$T/T_F(n)$ in a complicated way. However, for a gas expanding
under isentropic conditions, an exact scaling solution can be
obtained, and it predicts temperature-independent expansion. The
results correspond closely with our measurements, where nearly
temperature-independent  breathing frequencies are observed, as we
now show.

If the gas is locally isentropic, according to
Eq.~\ref{eq:localentropy}, the local reduced temperature does not
change as the gas expands, i.e.,
$T(\mathbf{x})/T_F[n(\mathbf{x})]=T_0/T_F[n_0(\mathbf{\tilde{x}})]$.
Using $n=n_0/\Gamma$ then requires
$T(\mathbf{x})=T_0/\Gamma^{2/3}$. If local equilibrium is
maintained, the pressure $P$ is then simply related to $P_0$.
Using
$f_P[T/T_F[n(\mathbf{x})]]=f_P[T_0/T_F[n_0(\mathbf{\tilde{x}})]]$
we must have $P(\mathbf{x})=P_0(\mathbf{\tilde{x}})/\Gamma^{5/3}$.
Then with $x =b_x\tilde{x}$ and
$\nabla_\mathbf{\tilde{x}}\rightarrow\nabla_\mathbf{x}$ in
Eq.~\ref{eq:forcebal2} for $P_0(\mathbf{\tilde{x}})$, we obtain
$\nabla P(\mathbf{x})/n(\mathbf{x})=-\nabla
U(\mathbf{\tilde{x}})/\Gamma^{2/3}$. Using this result in
Eq.~\ref{eq:hydro} yields
\begin{equation}
m\frac{\partial\mathbf{u}}{\partial
t}=-\nabla\left(\frac{m}{2}\mathbf{u}^2\,+\,U(\mathbf{x})-\frac{U(\mathbf{\tilde{x}})}{\Gamma^{2/3}}\right),
\label{eq:isentropichydro}
\end{equation}
where $U(\mathbf{x})$ is the trapping potential while
$U(\mathbf{\tilde{x}})$ arises from the pressure force, which is
evaluated at $\tilde{x}=x/b_x(t)$ and similarly for
$\tilde{y},\tilde{z}$. Eq.~\ref{eq:isentropichydro} can be
obtained  for a strongly collisional gas by using a phase-space
scaling ansatz~\cite{Pedri}.

We see that under isentropic conditions, the gas obeys the same
hydrodynamic equation at all temperatures as for a zero
temperature gas, where $P=c\,n^{5/3}$ and $c$ is a constant, even
though the pressure and density may assume a more complicated
form. For a harmonic potential, all terms in
Eq.~\ref{eq:isentropichydro} are linear in $x$ and the scaling
solution is exact~\cite{Menotti}. Hence, after release from a
harmonic trap, where  $U(\mathbf{x})\rightarrow 0$ in
Eq.~\ref{eq:isentropichydro}, the cloud expands precisely by a
scale transformation at all temperatures, similar to that observed
in our previous measurements for a strongly-interacting Fermi gas
at very low temperature~\cite{OHaraScience}.

The breathing frequencies  are readily obtained using
$b_i=1+\epsilon_i$, where $\epsilon_i <<1$. For the radial mode in
a cylindrically-symmetrically  trap, i.e., with
$\omega_x=\omega_y=\omega_\perp>>\omega_z$, one obtains
$\omega=\sqrt{10/3}\,\omega_\perp=1.83\,\omega_\perp$. However,
for our trap conditions, the exact result is  $\omega
=1.84\,\omega_\perp$, independent of the mean field and superfluid
contributions which are included in the general form of the energy
density.
\begin{figure}
\includegraphics[width=3.5in]{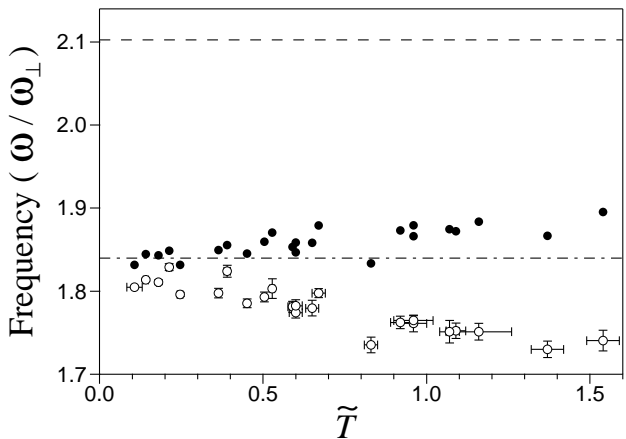}
 \caption{Radial breathing frequency $\omega/\omega_\perp$ versus empirical temperature
 $\tilde{T}$ for a unitary gas of $^6$Li from Ref.~\cite{KinastDampTemp}, showing small variation. Open circles:
 Measured frequencies; Solid dots: Data after correction for anharmonicity;
 Lower dot-dashed line: unitary hydrodynamic frequency, $\omega/\omega_\perp=1.84$.
 Upper dashed line: noninteracting gas frequency $2\omega_x/\omega_\perp =2.10$.} \label{fig:freq}
\end{figure}

Fig.~\ref{fig:freq} shows the measured breathing mode frequencies
as a function of empirical temperature $\tilde{T}$,
$0.11\leq\tilde{T}\leq 1.54$. Here, $\tilde{T}$ is determined from
the measured spatial profiles as in Ref.~\cite{JointScience} and
can be calibrated to the theoretical profiles~\cite{JointScience},
which show that the corresponding $T/T_F$ varies from $0.12$ to
$1.1$. After correction for anharmonicity, the frequency is close
to the universal hydrodynamic value over the range of temperatures
studied. These results are consistent with nearly isentropic
conditions at the highest temperatures, although the system is
likely to be changing from a superfluid to a unitary collisional
fluid, as suggested by the transition in the damping
rate~\cite{KinastDampTemp}. The observed hydrodynamic behavior at
higher temperatures is not explained by existing theories. In
particular, the momentum relaxation rate predicted in two-body
collision models is much too small to explain the observed
hydrodynamic behavior~\cite{BruunViscous}.

We are indebted to Jason Ho for a discussion of isentropic
hydrodynamics, and to Jason Ho, Kathy Levin and Qijin Chen for a
critical reading of the manuscript. This research is supported by
the Physics Divisions of the Army Research Office and the National
Science Foundation, the Physics for Exploration program of the
National Aeronautics and Space Administration, and the Chemical
Sciences, Geosciences and Biosciences Division of the Office of
Basic Energy Sciences, Office of Science, U. S. Department of
Energy.


\begin{thebibliography}{29}
\bibitem[{bar()}]{baremass}
\bibinfo{note}{We use the bare mass instead of an effective
  mass~\cite{Baker,JointScience}, since interactions are already generally
  included in the pressure.}
\end{thebibliography}

\end{document}